\documentclass[12pt,preprint]{aastex}
\shortauthors{C.G. Mundell et al.}  

\shorttitle{Sub-kiloparsec \HI disks in Arp~220}

\begin{document}

\title{Nuclear gas dynamics in Arp~220 $-$
sub-kiloparsec scale atomic hydrogen disks}

\author{C.G. Mundell\altaffilmark{1,2,3}}
\email{cgm@astro.umd.edu}
\altaffiltext{1}{Astrophysics Research Institute, Liverpool John Moores
University, Twelve Quays House, Egerton Wharf, Birkenhead, CH41 1lD, UK}
\altaffiltext{2}{Department of Astronomy, University of Maryland,
College Park, MD20742, USA}
\altaffiltext{3}{Royal Society University Research Fellow}

\author{P. Ferruit\altaffilmark{2,4}}
\email{ferruit@cumulus.univ-lyon1.fr}
\altaffiltext{4}{Centre de Recherche Astronomique de Lyon, 9
av. Charles Andr\'e, 69561 Saint-Genis Laval Cedex, France}

\author{A. Pedlar\altaffilmark{5}}
\email{ap@jb.man.ac.uk}
\altaffiltext{5}{University of Manchester, Jodrell Bank Observatory,
Macclesfield, Cheshire, SK11 9DL, UK}

\def\fsec{\hbox{$.\!\!^{s}$}}
\def\farcs{\hbox{$.\!\!^{\prime\prime}$}}
\def\HI{\hbox{H\,{\sc i}}}
\def\cmsq{cm$^{-2}$}
\def\NH{N$_{\mbox{\scriptsize H}}$}
\def\kms{km~s$^{-1}$}
\def\mJyb{mJy beam$^{-1}$}
\def\fdeg{\hbox{$.\!\!^{\circ}$}}

\begin{abstract}
We present new, high angular resolution ($\sim$0$\farcs$22) MERLIN
observations of neutral hydrogen (\HI) absorption and $\lambda$21-cm
radio continuum emission across the central $\sim$900 parsecs of the
ultraluminous infrared galaxy, Arp220. Spatially resolved \HI\
absorption is detected against the morphologically complex and
extended $\lambda$21-cm radio continuum emission, consistent with two
counterrotating disks of neutral hydrogen, with a small bridge of gas
connecting the two. Column densities across the two nuclei are high,
lying in the range 8$\times$10$^{19}$~T$_{\rm
s}$(K)$\lesssim$~\NH~$\lesssim$${\rm 2.4}
\times {\rm 10}^{20}$~T$_{\rm s}$(K)~\cmsq\ (T$_{\rm S}$ is spin
temperature) and corresponding to optical extinctions of
0.052~T$_{\rm s}$(K)$\lesssim$~A$_V$~$\lesssim$~0.155~T$_{\rm
s}$(K)~mag, with the higher column densities in the eastern nucleus.
Velocity gradients are clearly visible across each nucleus, reaching
1010 $\pm$ 20~\kms\,kpc$^{-1}$ in PA $\sim$ 55\arcdeg\ and 830 $\pm$
20~\kms\,kpc$^{-1}$ in PA $\sim$270\arcdeg\ for eastern and western
nuclei respectively. These gradients imply dynamical masses
M$_D$~=~1.1~$\times$~10$^{9}$~(sin$^{-2}i$)~M$_{\odot}$~(E) and
1.7~$\times$~10$^{8}$~(sin$^{-2}i$)~M$_{\odot}$~(W), assuming the
neutral gas is distributed in two thin circularly rotating disks.

We propose a merger model in which the two nuclei represent the galaxy
cores which have survived the initial encounter and are now in the
final stages of merging, similar to conclusions drawn from previous CO
studies (Sakamoto, Scoville \& Yun 1999 - SSY99 hereafter). However,
we suggest that instead of being coplanar with the main CO disk (in
which the eastern nucleus is embedded), the western nucleus lies above
it and, as suggested by bridge of \HI\ connecting the two nuclei, will
soon complete its final merger with the main disk. We suggest that the
collection of radio supernovae (RSN) detected in VLBA studies in the
more compact western nucleus represent the second burst of star
formation associated with this final merger stage and that free-free
absorption due to ionised gas in the bulge-like component can account
for the observed RSN distribution.

\end{abstract}

\keywords{galaxies: individual (Arp~220); galaxies: kinematics and
      dynamics; galaxies: nuclei}

\section{Introduction}

Ultra-Luminous Infra-Red Galaxies (ULIRGs), identified by IRAS and
defined by their high far-infrared luminosities ($>$10$^{12}$
L$_\odot$), have spectral energy distributions dominated by the
8-1000-$\mu$m mid-far IR emission (e.g., Soifer et al. 1987; Clements
et al. 1996a; Sanders \& Mirabel 1996) and show luminosities and space
densities in the local Universe similar to those of QSOs (Soifer 1986;
Soifer, Houck \& Neugebauer 1987). The excess of far-IR emission in
these objects is generally thought to indicate the presence of large
amounts of hot dust, but the mechanism for heating the dust remains
controversial. Intense nuclear starbursts and heavily-shrouded active
galactic nuclei (AGN) have both been advocated and a combination of
both energy sources currently seems likely, with starbursts dominating
at L$_{\rm IR}$$<$10$^{12.3}$~L$_{\odot}$ and the AGN-dominated
fraction increasing to 35\%$-$50\% above this luminosity (Genzel et
al. 1998; Veilleux, Sanders \& Kim 1999).

ULIRGs are almost always found in strongly interacting systems or
advanced mergers (Clements et al. 1996a,b; Clements \& Baker 1996),
during which large-scale torques on the intersteller medium and loss
of angular momentum drives gas towards the central few kilo-parsecs of
the system, where it is thought to result in intense starburst and/or
AGN activity (e.g., Barnes \& Hernquist 1991; Mihos \& Herquist 1994,
1996; Norman \& Scoville 1988). In this scenario, an (obscured) AGN is
switched on as the galaxy merger reaches an advanced state, and is
then progressively unveiled as the circumnuclear starburst processes
(and expels) the surrounding gas. Obscuration is high in the mid-IR
emitting regions of ULIRGs, with model-dependent extinctions ranging
from A$_{\rm v}$$\sim$5$-$1000 (Genzel et al. 1998). However,
progression from starburst- to AGN-powered emission as the
merger advances is not evident in present studies (e.g., Genzel et
al. 1998). If ULIRGs do contain obscured progenitors of future
optically-bright quasars (e.g. Sanders et al. 1988) they may offer a
valuable opportunity to study an important stage in the evolution of
quasars (and powerful radio galaxies) in the local
Universe. Alternatively, highly obscured AGN might not be common in
the ULIRG population since strong AGN activity, once triggered, might
quickly destroy the obscuring screen, producing an easily detectable
AGN (e.g. Lutz, Veilleux \& Genzel 1999).

In this context, the prototypical ULIRG Arp~220 (IC4553/4, UGC09913),
a double-nucleus system, is an extremely interesting object. In the
diagnostic diagrams of Genzel et al. (1998, their Fig.~5) it appears
as one of the most completely starburst-powered ULIRG in their sample,
despite the fact that it is in an advanced merger state. In these
final stages of merging, peak gas densities (and related activity) are
thought to occur, when the merging galaxies are within a $\sim$kpc of
one another and are in the process of coalescing (Mihos
\& Hernquist 1996); high angular resolution observations of
the central kiloparsec are therefore vital to determine the gas
dynamics in this important region. This paper presents the first
subarcsecond-resolution, spatialling-resolved spectral imaging of
the distribution and kinematics of neutral hydrogen (\HI) in the
central 900 pc of Arp~200, improving on previous observations by more
than a factor of ten in angular resolution.

HI is a valuable tracer of galactic structure and dynamics over a wide
range of size scales and is a valuable tool for probing probe the
gaseous conditions in the highly-obscured, compact nuclear regions of
ULIRGs which are often unaccessible at optical wavelengths.  Indeed
ULIRGs contain large amounts of neutral hydrogen (e.g. \NH\ $>$
10$^{21}$$-$10$^{22}$ atoms
\cmsq\ - Mirabel 1982), and recent VLA synthesis maps of \HI\ {\it
emission} in objects such as Arp~220 (Hibbard, Vacca, \& Yun 2000),
have revealed massive gaseous tidal features (M$_{\rm H}$ $>$ 10$^9$
M$_{\odot}$) extending over several tens of kpc beyond the optical
emission and demonstrating the dramatic redistribution of material
during galaxy merger.  Unfortunately, such studies of \HI\ in {\em
emission} are limited by surface brightness sensitivity to angular
resolutions of $>$6$''$$-$10$''$ with current instruments. Instead,
\HI\ {\em absorption} measurements can be performed to angular
resolutions as high as a few milliarseconds in the presence of a
bright radio continuum background source (e.g. Peck
\& Taylor 1998; Carilli \& Taylor  2000); subarcsecond
resolutions achievable with MERLIN probe the distribution and
kinematics of \HI, in absorption, on scales of $\sim$10$-$100 pc in
nearby galaxies (e.g. Mundell et al. 1995; Cole et al. 1998). 

Here we present new, 0\farcs22 (81-pc) resolution, $\lambda$21-cm
\HI\ absorption observations of the central $\sim$900~pc of Arp~220 with
MERLIN.  The observations and data analysis are described in
Sect.~\ref{SectionObservations} and the results in
Sect.~\ref{SectionResults}. We then compare the derived masses with
those obtained from the molecular gas observations, and discuss the
present geometry of the system and the history of the merger in view
of these new observations and the available data. We adopt a distance
of $\sim$ 77~Mpc for Arp~220, at which 1\arcsec\ correspond to $\sim$
370~pc.

\section{Observations and Data Analysis}
\label{SectionObservations}
   
   \subsection{Observations and data calibration}
      \label{SectionObservationsReduction}
      
The central region of Arp~220 was observed at $\lambda$21~cm, on 26th
January, 1996, with the Multi-Element Radio Linked Interferometer
(MERLIN), which at this time consisted of 8 telescopes (including the
76-m Lovell telescope). The maximum baseline length of the array is
217~km, corresponding to $\sim$ 1~M$\lambda$ at $\lambda$21~cm and
yielding an effective spatial resolution of 0\farcs22. Dual
polarizations were recorded for an 8-MHz bandwidth in spectral line
mode, with 64 channels per polarization, resulting in a channel width
of 125 kHz (26.3~\kms). As the first channel is reserved for on-line
information, the centre of the band lies at channel 31.5, corresponding
to an optical heliocentric velocity of 5496.2~\kms. Total observing
time on Arp~220 was $\sim$7 hours, with 2 minutes observations of the
phase calibrator 1551+239 interleaved every 5 minutes throughout the
observations. Absolute flux calibration was determined from
observations of 3C286, assuming a flux density of 14.731 Jy at 1420.4
MHz (Baars et al. 1977). 0552+398 was used to determine bandpass
corrections for the spectral line data.

After preliminary amplitude calibration and gain-elevation corrections
were applied, using local MERLIN software, the data were transferred
to the NRAO Astronomical Image Processing System ({\sc aips} - van
Moorsel, Kemball \& Greisen 1996) for all subsequent calibration,
editing and imaging. Standard calibration and data editing (Greisen \&
Murphy 1998) were performed on the `pseudo-continuum' dataset, which
is formed by averaging the central 75\% (6 MHz) of the band (analogous
to the VLA `channel 0' dataset). The resulting map of the phase
calibrator was used to guide additional data editing and
self-calibration; the derived phase and gain corrections were applied
to the Arp~220 pseudo-continuum data, from which a preliminary image
was produced and used in further data editing and phase
self-calibration. The total corrections and data flags were then
applied to the spectral line data and bandpass calibration was
performed. Due to good bandpass stability, no significant improvement
to the continuum images was achieved by iteratively performing
self-calibation and bandpass calibrations. Finally, the telescopes
were reweighted according to their gains.
    
The continuum contribution (formed from the line-free channels 5-15
and 50-60) was subtracted in the $(u,v)$ plane using {\sc aips} task
{\sc uvlin}. The corresponding `true' continuum dataset was formed
from the same line-free channels in the $(u,v)$ plane using {\sc
avspc}. These two datasets were then Fourier transformed with uniform
weighting (pixel size 45~mas), deconvolved and added together to form
the absorption cube (512~$\times$~512~$\times$~63) with the correct
continuum level adjacent to the line. A circular restoring beam of
size 0\farcs22 was used for both datasets and the r.m.s. noise
levels were $\sim$ 0.1~\mJyb\ and 0.5 \mJyb\ per channel for the
continuum image and spectral line cube, respectively.

\subsubsection{Positional Accuracies}
\label{SectionObservationsPositions}
      
The commonly-used phase-referencing technique of interspersing
observations of the target source with regular and frequent
observations of a nearby calibrator with known position allows the
absolute position of the target source to be determined; the accuracy
of the calibrator position is therefore important.  Our phase
calibrator, 1551+239, was originally identified in the calibrator
survey of Patnaik et al. (Patnaik et al. 1992; Browne et al. 1998) and
lies in `Region 2' or declination range 20$-$35\arcdeg\ (Wilkinson et
al. 1998). The typical positional uncertainty for the survey is 50 mas
but `Region 2' suffered from larger systematic errors than other
regions of the survey. The origin of these errors is unknown and the
positional uncertainty for 1551+239, which was omitted from the final
published survey, might be as large as 100 mas, predominantly in
declination (Thomasson, private communication).  Positions derived
from previous VLA images at 5 GHz (Baan \& Haschick (1995), 15 and 22
GHz (Norris, 1988) also show a range of values, differing primarily in
declination by $\sim$100$-$200 mas. Consequently, the positional
uncertainty for Arp~220 is no better than 100 mas.

Nevertheless, the derived position for the western nucleus, 
$\alpha_{\rm B1950}$ = 15$^h$32$^m$46\fsec89142s,  $\delta_{\rm B1950}$
= 23\arcdeg40\arcmin 08\farcs0215, compares well with positions
derived from CO observations by SSY99, $\alpha_{\rm 1950}$ =
15$^h$32$^m$46\fsec88s, $\delta_{\rm 1950}$ = 23\arcdeg40\arcmin
08\farcs0 ($\pm$0\farcs1), and 5-GHZ MERLIN observations by Baan
(private communication), $\alpha_{\rm 1950}$ =
15$^h$32$^m$46\fsec89053s, $\delta_{\rm 1950}$ = 23\arcdeg40\arcmin
08\farcs0509.  It should be noted that although the positional
uncertainties limit very accurate absolute registration with
observations from other instruments, the MERLIN results presented here
are not otherwise affected since Arp~220 is strong enough for
self-calibration to be performed.

\subsection{Analysis of the \HI\ absorption profiles}
      \label{SectionObservationsAnalysis}
      
Examination of the observed \HI\ absorption line profiles revealed
that not all are well approximated by a single Gaussian profile, due
to the presence of additional narrow components and/or wings
overimposed on the main, relatively broad component. Simple, automatic
moment analysis is therefore difficult and may result in corrupted
velocity fields, where a single Gaussian component does not represent
the centroid velocity accurately. We therefore performed a multiple
Gaussian profile fitting of the absorption spectra using a modified
version of the {\sc fit/spec} software (Rousset 1992), originally
developed to fit optical emission lines.
      
The fitting was performed in two stages. First a single Gaussian
profile was fitted automatically to each absorption spectrum and
residuals were examined to detect regions where additional components
where likely to be present (i.e.  regions with residuals higher than
three times the $\sigma$ = 0.5~\mJyb\ noise level) - these regions
coincided with each nucleus. For each nucleus, the spectral
characteristics of the additional components were then determined by
fitting to individual spectra with the most contrast.  Next, a
multi-component Gaussian fit to the line profiles was performed, using
one relatively broad Gaussian component (the `main' component) and a
set of narrower components with fixed characteristics (see Table
1). This second step was successful for the western nucleus which
exhibited the strongest, most distinct, additional components, but
failed for the eastern nucleus where the additional components were
poorly contrasted (peak amplitude typically lower than 6$\sigma$).
Therefore, the moment maps presented in this paper correspond to a
single Gaussian component fit to spectra from off-nuclear regions and
the eastern nucleus, and a multiple Gaussian component fit to spectra
from the western nucleus. Errors on the centroid velocity in the
eastern nucleus due to the use of a single Gaussian profile are
typically lower than 10~\kms.

\section{Results}
\label{SectionResults}

\subsection{$\lambda$21-cm  radio continuum emission}
   \label{SectionResultsContinuum}


The MERLIN image of the $\lambda$21-cm (1.4-GHz) radio continuum
emission from the nuclear region of Arp~220 is shown in
Figure 1. The two nuclei (Baan et al.  1987; Norris 1988; Baan
\& Haschick 1995, BH95 hereafter) are clearly resolved and are
embedded in more complex extended emission. The nuclei are labelled E
and W in Figure1 (labelled A and B in the 5-GHz image of
BH95).  Their peak and integrated fluxes\footnote{The uncertainty in
the flux scale is taken to be 5\% and is included in the total
uncertainties in flux densities; these errors were derived by adding,
in quadrature, the 5\% amplitude error, the rms noise in the image and
the error in the Gaussian fitting.} are 28.7$\pm$1.4~\mJyb\
\& 126.5$\pm$6.4~mJy (E) and 64.9$\pm$3.3~\mJyb\ \& 129.3$\pm$6.5~mJy
(W), respectively.  The total integrated flux density of the source is
$\sim$ 285.4$\pm$14.3~mJy. A large-area box flux yields a slightly
higher integrated flux, indicating the presence of structures extended
on scales larger than 1\arcsec-2\arcsec. The presence of extended
emission is also suggested by the higher integrated flux densities of
302~mJy (White \& Becker, 1992) and 312~mJy (Condon \& Dressel, 1978)
measured with single dish telescopes at 1.4 GHz and 2.4 GHz
respectively. Gaussian fitting to the MERLIN image reveals that the
eastern nucleus is well resolved (as can be seen in Figure 1), with a
deconvolved size (FWHM) of 440~$\times$~345~mas (i.e. $\sim$160
$\times$ 130~pc) in PA = 55\fdeg0~$\pm$~0\fdeg7. The western nucleus
is only marginally resolved with a deconvolved size (FWHM) of
245~$\times$~190~mas (i.e. 90 $\times$ 70~pc) in PA =
96.3\arcdeg~$\pm$~0\fdeg7. The derived position for the western
nucleus is $\alpha_{\rm 1950}$ = 15$^h$32$^m$46\fsec89142s,
$\delta_{\rm 1950}$ = 23\arcdeg40\arcmin 08\farcs0215 which
corresponds well with observations at other frequencies (see Section
\ref{SectionObservationsPositions} for full discussion).
   
In addition to emission from the two nuclei, a tongue of
$\lambda$21-cm continuum emission is seen to the south-east of the
western nucleus (Figure 1) and corresponds to emission previously
identified by BH95 (component C in their paper) at 5~GHz. Similar weak
extensions are seen around the eastern nucleus. Finally, a broad spur
of continuum emission (labelled T in Figure 1) is seen north-west of
the western nucleus. It extends over 0\farcs8 in PA $\sim
-$50\arcdeg. This weak component is detected over $\sim$ 11 beam
areas, with a peak flux density of $\sim$ 1.6~\mJyb\ and an integrated
flux density of 6.6~\mJyb. There is a hint for the presence of a
similar component in the 5-GHz naturally weighted image of BH95 (their
Fig.~1, upper panel), but their angular resolution (0\farcs43) is
insufficient to deblend it from the emission of the western nucleus.
   
\subsection{Neutral Hydrogen Absorption}
   \label{SectionResultsAbsorption}

   
Figure 2 shows the MERLIN spectral channel images of \HI\
absorption detected across the central $\sim$600~pc of
Arp~220. Comparison of the channel images with the continuum image,
which is displayed to the same scale in the top left panel, shows that
absorption is clearly detected against each continuum nucleus, as well
as against the bridge of continuum that connects the two nuclei.
Figure 3 shows some representative absorption spectra,
taken from eight locations across the continuum structure; many of the
absorption lines are asymmetric with some showing additional velocity
components (e.g. location `ER').

The ability to image neutral gas in absorption depends on both the
structure of the background continuum emission and the distribution of
the foreground absorbing gas; the neutral gas in Arp~220 is spatially
extended and so the absorption is well resolved at this angular
resolution, in particular against the extended continuum emission of
the eastern nucleus. The velocity range over which absorption is
detected ranges from $\sim$ 5160~\kms\ to 5725~\kms\ and compares well
with the width of the single dish absorption line at 10\% of the peak
absorption (Mirabel \& Sanders 1988). However, this is lower than the
full width at zero intensity (FWZI) of the single dish profile, $\sim$
640~\kms\ (Mirabel \& Sanders 1988). If weak absorption is present
against weak, extended radio continuum emission, this absorption would
not be detectable in the MERLIN observations, and this may explain the
additional 50-\kms\ width at the red and blue extremes of the single
dish profile.

H{\sc i} absorption is not detected against the continuum component C, to
the south-east of the western nucleus (Figure 3), to a
limiting 3$\sigma$ absorption depth of 1.5~\mJyb. Assuming a constant
column density across C, this implies a maximum column density
\NH~$\sim$ 1.7 $\times$ 10$^{19}$ T$_{\rm s}$(K)~\cmsq, where T$_{\rm
s}$(K) is the spin temperature for which the exact value is unkown but
is found to be typically $\sim$10$^2$$-$10$^4$ K in the Galaxy (Heiles
\& Kulkarni 1988). The weakness of continuum component C prevents the
determination of stringent upper limits for \NH, but the relatively
blue colour of the infrared component coincident with C (Scoville et
al. 1998) provides support for a low column density in this region.

Very weak \HI\ absorption may be present against the spur of weak,
extended continuum emission (T) to the north-west of the western
nucleus, but the continuum emission is weak in this region (peak flux
density $\sim$ 1.6~\mJyb), preventing a reliable determination of
column density; a 3$\sigma$ detection would require a column density
\NH~$>$~1.3~$\times$~10$^{20}$~T$_{\rm s}$(K)~\cmsq. Close inspection
of the spectral channel images in Figure 2, over a velocity
range $\sim$5500 $-$ 5300~\kms, suggests a possible weak absorption
component extended in the same PA as the continuum extension ($\sim
-$50\arcdeg), but more sensitive observations would be required to
confirm its detection and  measure the column density.

Figure 4 shows the two-dimensional column density distribution and
velocity field of the main kinematic component, as derived from
multi-component Gaussian profile fitting of the spectra (see
Sect.~\ref{SectionObservationsAnalysis}). Characteristics of the main
and additional kinematic components in the eight spectra displayed in
Figure 3 are listed in Table 1 and examples of
Gaussian fitting to components in the eastern and western disks are
shown in Figure 5 and Figure 6, respectively. The two nuclei are
spatially well separated at this resolution (as shown in
Figure 2) and a velocity gradient is clearly visible
across each nucleus. Across the eastern nucleus, the velocity gradient
is 1010 $\pm$ 20~\kms\,kpc$^{-1}$ in PA $\sim$ 55\arcdeg, while the
marginally resolved western nucleus shows a gradient of 830 $\pm$
20~\kms\,kpc$^{-1}$ in PA $\sim$270\arcdeg.
   
\subsubsection{Nuclear Column Densities}

The neutral hydrogen column densities across the two nuclei are high,
lying in the range 8$\times$10$^{19}$~T$_{\rm s}$(K)$\lesssim$ \NH
$\lesssim$${\rm 2.4}\times{\rm 10}^{20}$~T$_{\rm s}$(K)~\cmsq , with
the higher column densities generally being found in the eastern
nucleus.  The corresponding optical extinctions, if \HI\ column
density and optical extinction in Arp~220 follow the same relation as
that seen in the Galaxy (Staveley-Smith \& Davies 1987), are
0.052~T$_{\rm s}$(K)$\lesssim$ A$_V$ $\lesssim$ 0.155 T$_{\rm s}$(K)
mag and are detailed in Tab.~\ref{TableResultsFit}.  The \HI\ spin
temperature is unknown, but for the minimum \HI\ spin temperature
T$_{\rm S}$$\sim$100-200 K, these extinction values compare well with
those derived by Scoville et al. (1998) from high resolution infrared
{\em HST} NICMOS images at 1.1, 1.6 and 2.22 $\mu$m and confirm that
the radio continuum nuclei lie in regions of relatively high
extinction.

An estimate of the column density of molecular hydrogen can be
inferred from the CO observations of SSY99, who measured integrated CO
fluxes S$_{\rm CO(2-1)}$ = 120 Jy \kms\ (E) and 187 Jy \kms\
(W). Assuming an excitation temperature of 40~K, a CO:H$_{\rm 2}$
abundance X(CO)~=~2.7~$\times$~10$^{-4}$ (Lacy et al. 1994) and
optically thin emission, we infer H$_{\rm 2}$ column densities N$_{\rm
H_2}$$\sim$2.8~$\times$~10$^{22}$ \cmsq\ (E) and 4.3~$\times$~10$^{22}$
\cmsq\ (W). These values are
comparable to the mean \HI\ column densities \NH\ = 1.6 $\times$
10$^{20}$ T$_{\rm s}$ \cmsq\ (E) and 1.3 $\times$ 10$^{20}$ T$_{\rm
s}$ \cmsq\ (W) for T$_{\rm S}$$\sim$100$-$200 K and result in total
hydrogen column densities N(H)~$\sim$4.4 $\times$ 10$^{22}$ \cmsq\ (E)
and 5.6 $\times$ 10$^{22}$ \cmsq\ (W); these columns densities are
likely to be lower limits due to the uncertainty in the values of
T$_{\rm S}$ (Heiles \& Kulkarni 1988) and X(CO) (Frerking, Langer
\& Wilson 1982), excitation conditions and optical depth.

Indeed, the nature of the IR emission in Arp~220, AGN or
starburst-driven, has long been controversial and the amount of
obscuration present is important. High frequency observations have
failed to find a hard X-ray or soft $\gamma$-ray component consistent
with an AGN, suggesting that the column density required to obscure
any hidden AGN must be \NH\ $\gtrsim$10$^{25}$
\cmsq (Rieke 1988; Dermer et al. 1997). Alternatively, if \NH\
$\lesssim$ 10$^{24}$ \cmsq, more than 80\% of the total IR luminosity
must have a non-AGN origin.  Even if the \HI\ spin temperature is as
high as T$_{\rm s}$ $\sim$ 10$^{4}$ K, as might be expected in the
vicinity of strong radio continuum emission, the peak measured column
density is \NH\ $\sim$2.4~ $\times$10$^{24}$~\cmsq, suggesting that
either a large additional quantity of ionized and molecular gas is
required to provide the necessary obscuration towards any AGN
component or the IR emission is dominated by a starburst.

\section{Discussion}
  \label{SectionDiscussion}
   
  \subsection{The nuclear disks}
      
The kinematics of the main absorption component are consistent with
two counterrotating disks associated with the two nuclei. We therefore
confirm the results from the lower spatial resolution (0\farcs5) CO
observations of Sakamoto et al. (1999). In
particular, the velocity field of the disk in the more compact western
nucleus is clearly apparent in our higher resolution (0\farcs22)
observations. The kinematic major axis for this western nuclear disk
($\sim$ 270\arcdeg) is consistent with the value inferred from the CO
observations (263\arcdeg, SSY99, their Tab.~1) but is perpendicular to
the north-south velocity gradient inferred from the formaldehyde
emission observations of BH95.
      
This discrepancy can be explained if the gaseous component responsible
of the formaldehyde emission is distinct from the western disk
gas. The location of the peak of formaldehyde emission, $\sim$
0\farcs1 south of the western radio continuum peak (BH95), is
coincident with the position of the group of luminous radio supernovae
detected by Smith et al. (1998). This suggests that the source of the
pumping photons is the intense starburst occuring in the western disk
and therefore belongs to it. However, the masing gas can be located
far away from the continuum source. Baan, G\"usten \& Haschick (1986)
give {\em upper limits} on the distance between the continuum source
and the masing gas in Arp~220 ranging from 250~pc to 1.1~kpc. Given
these upper limits, the masing gas could be molecular gas located in
front of the western nucleus and unrelated to the western nuclear disk
or a warp in the eastern/large-scale molecular disk .  Indeed, the
horizontal isovelocity contours of CO(2-1) emission seen to the south
of the western nucleus (SSY99 - Fig.~3, upper right panel; Downes \&
Solomon 1998 - Fig.~18, upper right panel) would support this
interpretation for a non-coplanar component of molecular gas.
      
Using mean column densities \NH\ = 1.6 $\times$ 10$^{20}$ T$_{\rm s}$
\cmsq\ and 1.3 $\times$ 10$^{20}$ T$_{\rm s}$ \cmsq\ for the eastern
and western nuclei respectively and semi-major axes
167~pc~$\times$~118~pc (E) and 102~pc~$\times$~65~pc~(W), corresponding
to the radius at which the \HI\ absorption line strength is greater than
3$\sigma$, we infer \HI\ masses M$_{\rm H}$~=~7.5~$\times$ 10$^4$
T$_{\rm s}$ M$_{\odot}$ (E) and 2.0 $\times$ 10$^4$~T$_{\rm
s}$~M$_{\odot}$~(W). These masses are (1.3$-$5.1)$\times$ 10$^4$
T$_{\rm S}^{-1}$ times lower than the mass of molecular gas in the
disks derived from the CO observations ($\sim$ 10$^{9}$ M$_{\odot}$;
Downes
\& Solomon 1998, DS98 hereafter; SSY99); given the uncertainty in the
CO mass determination, the \HI\ and CO masses would be comparable for
high T$_{\rm S}$~$\sim$10$^4$~K. Assuming the neutral gas is
distributed in two circularly rotating disks of radius 167~pc (E) and
102~pc (W), the observed velocity gradients (see
Sect.~\ref{SectionResultsAbsorption}) imply dynamical masses
M$_D$~=~1.1~$\times$~10$^{9}$~(sin$^{-2}i$)~M$_{\odot}$~(E) and
1.7~$\times$~10$^{8}$~(sin$^{-2}i$)~M$_{\odot}$~(W).  The \HI-inferred
dynamical mass for E is slightly lower than values derived from CO
observations (2.4 and $>$ 1.9 $\times$ 10$^{9}$~(sin$^{-2}i$)
M$_{\odot}$ for DS98 and SSY99, respectively) and indeed the CO
emission (DS98, SSY99) out to a radius of $\sim$0\farcs3 (111 pc),
although weak and not fully resolved, appears to show a steeper
velocity gradient than that of the \HI. For the western nucleus, our
dynamical mass is very similar to that inferred by DS98
(1.7~$\times$~10$^{8}$~sin$^{-2}i$~M$_{\odot}$) but 10 times lower
than that inferred by SSY99
($>$~1.5~$\times$~10$^{9}$~sin$^{-2}i$~M$_{\odot}$) for a similar disk
radius. The reason for this difference is not clear and higher angular
resolution CO observations might determine whether the CO is located
deeper in the nuclear potential well than the HI.

   \subsection{Current geometry and history of the merger}
   \label{SectionDiscussionHistory}
      
The fact that the two nuclei have counterrotating gas disks, indicates
that the merger in Arp~220 involved a prograde-retrograde encounter of
two gas-rich progenitor disk galaxies. Numerical simulations of such
mergers can be found in Mihos \& Hernquist (1996, MH96 hereafter). In
particular, these simulations show that the prograde and retrograde
disks will behave differently during the merger and may also
experience different star formation episodes (MH96 - Sect.~4.2). In
Arp~220, qualitative comparison between the observations and the
models of MH96 suggests that the eastern nucleus, which rotates in the
same sense as the 1-kpc CO disk (Scoville et al. 1991; Scoville, Yun
\& Bryant 1997, SYB97 hereafter; DS98) and 100-kpc HI disk (Hibbard, Vacca
\& Yun 2000, HVY2000 hereafter) in which it is embedded,
represents the retrograde progenitor while the western nucleus and
north-western optical plume originate from the prograde progenitor
(HVY2000 hereafter).

Initially, during the first close passage, the retrograde disk (and
corresponding halo) sweeps up and accretes a significant fraction of
the gas of the prograde disk (30\,\% in the prograde-retrograde {\em
planar} encounter simulations of MH96).  In Arp~220, the eastern
nuclear disk as well as the kiloparsec-scale CO disk (Scoville et
al. 1991; SYB97; DS98), which comprise a large fraction of the mass of
the system, are probably the result of this process. Thus, given the
similarities between the PA and kinematics of these two overlaping
disks, it is likely that the two constitute a single (warped) disk,
and so we use the term `eastern disk' in the following discusision to
refer to the association of these two disks. The second argument in
favor of the eastern nucleus being the retrograde one, comes from the
morphology of the western nucleus. In the MH96 simulations and after
the first close passage, the prograde nucleus has lost up to 50\,\% of
its mass, either accreted by the retrograde disk or launched into
tidal tails. This is qualitatively consistent with the observed
properties of the western disk, which is more compact, less massive
than its companion and is likely the origin of the north-western
optical plume (HVY2000).
      
If the western nucleus is indeed the core of the prograde galaxy, this
has important implications for the current geometry of the
system. Based on the good agreement between the mean velocity of the
western disk and the line of sight velocity of gas in the large-scale
CO disk at this position (SYB97), previous studies have assumed that
the western disk was orbiting in the plane of the `large-scale' CO
disk (e.g. see Fig.~7 in SYB97 or Fig.~5 in SSY99). In this context,
the deprojected locations of the two nuclei were derived from their
mean line of sight velocities, and the western nucleus was placed
closer along our line of sight than the eastern nuclear disk (e.g. see
Fig.~24 in DS98).  However, for the western nucleus to be the prograde
nucleus its orbit must pass {\em behind} the eastern nucleus, so the
location of the western nucleus on its orbital path is therefore
important. The absence of an \HI\ absorption signature from
eastern-disk gas in the spectra of the western nucleus indicates that
the western disk lies above the eastern disk, while the mean
blueshifted velocity of the western disk with respect to the eastern
disk ( $\sim$ $-$170~\kms, SSY99), permits only the far-sided half of
the orbit (see Figure 7); combining these two constraints results in the western disk
lying on the orbit quadrant above the eastern disk, on its final
inward journey towards collision with the eastern disk (see Figure 7).
The presence of disturbed \HI\ in the bridge (B) connecting
the two nuclei further supports this merger scenario with the eastern
edge of the western disk merging with the main disk.
      
Continuing our comparison with the models of MH96 and assuming that
the eastern nucleus is the retrograde nucleus, we can investigate how
and when two important starburst events in Arp~220 fit in such a
scenario. The first event is that which gave birth to the galactic
superwind and the associated kiloparsec scale bubbles observed in
Arp~220 (Heckman, Armus \& Miley 1987 1990, HAM90 hereafter; Heckman
et al. 1996, HDEW96 hereafter). Given the kinematic expansion time of
the bubbles (3 $\times$ 10$^{7}$ yr, HDEW96, from HAM90) and the time
needed by a (continuous) starburst to deposit significant mechanical
energy into its environment ($<$ 10$^{7}$~yr, Fig.~56 in Leitherer \&
Heckman 1995, LH95 hereafter), the starburst powering this outflow
must have started a 10$^{7}$ to 10$^{8}$ years ago (additional time
may have been necessary for the superwind to break out from the
gaseous disk). As the orientation of the outflow is roughly
perpendicular to the major axis of the eastern disk, it is likely that
this starburst took place in this disk. Assuming that the Arp~220
merger is currently in the stage just before the final fusion of the
two gas rich systems (i.e. t $\sim$ 58-60 in Fig.~12 in MH96) this
corresponds\footnote{The absolute value of the time step in the
simulations of MH96 depends on the initial characteristics of the
galaxies. For values appropriate for the Milky Way, one time step
corresponds to t $\sim$ 1.3 $\times$ 10$^{7}$~yr (see MH96 -
Sect.~2.1).} to t $\sim$ 50-56 in the simulations. In the simulations,
the retrograde disk does indeed experience a continuous period of star
formation (which is stronger than in the prograde disk, MH96) during
these intermediate times (Fig.~13 in MH96). Note that, assuming that
all the mechanical energy deposited by the starburst is injected into
the wind, the mechanical luminosity of the superwind ($\sim$ 10$^{43}$
erg~s$^{-1}$, HDEW96) corresponds to a star formation rate of $\sim$
10-100~M$_{\odot}$~yr$^{-1}$ (LH96 - Fig.~56 for a continuous star
formation rate). These values are lower than the recent star formation
rate estimated in Arp~220 ($\sim$ 240~M$_{\odot}$~yr$^{-1}$,
Anantharamaiah et al.  2000), but are consistent with the models which
predict a sharp increase in the star formation rate in the final stage
of the merger compared to the intermediate times (MH96 - Fig.~14,
second row).
      
The second starburst event, more recent and probably more violent
(consistent with the models of MH96 described above), is the one
revealed by the large number of luminous radio supernovae detected in
the western nucleus (Smith et al. 1998, SLLD98 hereafter). The rate of
type II supernovea inferred from these observations is 1-4 yr$^{-1}$
and corresponds to a star formation rate ranging from 50 to
800~M$_{\odot}$~yr$^{-1}$ (LH95 - Fig.~6 for a continuous star
formation rate). This is consistent with the value of
160~M$_{\odot}$~yr$^{-1}$ that can be inferred\footnote{Roughly two
thirds of the H92$\alpha$ line emission in their observations come
from the western nucleus. If we scale the total star formation rate
(240~M$_{\odot}$~yr$^{-1}$) accordingly, this gives a star formation
rate for the western nucleus alone of $\sim$
160~M$_{\odot}$~yr$^{-1}$.} for the western nucleus from the
observations of Anantharamaiah et al. (2000). One striking feature of
the distribution is the luminous radio supernovae (SLLD98). Figure 8
shows the positions of the radio supernovae and OH megamaser emission
features with respect to the MERLIN 1.4-GHz continuum emission. The
fitted positions of the E and W nuclei as derived from the MERLIN
1.4-GHz continuum image (Sect. \ref{SectionResultsContinuum}) and from
similar MERLIN observations made at 5 GHz (Baan, private
communication) are also shown.  There is a clear offset between the
RSN and the nuclear positions, particularly for the western nucleus
where the supernovae are located $\sim$0\farcs1~$-$~0\farcs15 {\em
south} of the western nucleus and consequently, the compact central
feature of the W1 OH megamaser emission (Lonsdale et al. 1998a 1998b)
coincides with the position of the western nucleus.  In a simplistic
model of star formation occurring in an edge on disk, the RSN
distribution would be expected to coincide with the nucleus. However,
if star formation is occurring throughout the disk and the nearside of
the disk is tilted slightly to the south, as suggested by NICMOS
observations of Arp~220 (Scoville et al. 1998), RSN in the near half
of the disk would be more easily visible. In particular a moderate
column density of ionized gas associated with the bulge-like component
suggested by Scoville et al. (1998) and the longer path length to RSN
on the far side of the disk would result in RSN radio emission being
free-free absorbed.  Emission measures
1.1$\times$10$^{7}$$\lesssim$~EM~$\lesssim$2.4$\times$10$^{7}$~cm$^{-6}$~pc,
corresponding to optical depths 1.2$\lesssim$~$\tau$~$\lesssim$2.6,
would be required to free-free absorb emission from RSN with the same
flux density at 1.67 GHz as those observed by SLLD98, assuming a
uniform screen of ionized absorbing gas with an electron temperature
T$_{\rm e}$=10$^{4}$~K. Such emission measures compare well with those
inferred from radio recombination line observations (Anantharamaiah et
al. 2000) which, however, suggest more complex models for the ionized
gas than a single density uniform slab.  Alternatively, the RSN
positional offset might be a consequence of positional uncertainty of
the MERLIN observations (see Sect.
\ref{SectionObservationsPositions}); future observations are required
to confirm the registration of MERLIN and VLBA images.


\section{Conclusions}

We have presented the highest angular resolution observations to date
of the neutral gas distribution and kinematics and the $\lambda$21-cm
continuum emission in the central $\sim$900 pc of the advanced merger
system, Arp~220.  Spatially resolved \HI\ absorption was detected
against the morphologically complex and extended $\lambda$21-cm radio
continuum emission; the two nuclei are spatially well separated at
this resolution and a velocity gradient is clearly visible across each
nucleus consistent with the
\HI\ being in two counterrotating disks of neutral hydrogen, with a
small bridge of gas connecting the two. The column densities across
the two nuclei are high, lying in the range 8$\times$10$^{19}$~T$_{\rm
s}$(K)$\lesssim$~\NH~$\lesssim$${\rm 2.4}
\times {\rm 10}^{20}$~T$_{\rm s}$(K)~\cmsq\ (T$_{\rm S}$ is spin
temperature) and corresponding to optical extinctions of
0.052~T$_{\rm s}$(K)$\lesssim$~A$_V$~$\lesssim$~0.155~T$_{\rm
s}$(K)~mag, with the higher column densities being found in the
eastern nucleus. Velocity gradients of 1010 $\pm$ 20~\kms\,kpc$^{-1}$
in PA $\sim$ 55\arcdeg\ and 830 $\pm$ 20~\kms\,kpc$^{-1}$ in PA
$\sim$270\arcdeg\ across the eastern and western nuclei respectively
imply corresponding dynamical masses
M$_D$~=~1.1~$\times$~10$^{9}$~(sin$^{-2}i$)~M$_{\odot}$~(E) and
1.7~$\times$~10$^{8}$~(sin$^{-2}i$)~M$_{\odot}$~(W), assuming the
neutral gas is distributed in two thin, circularly rotating disks.  

We propose a new merger geometry in which the two nuclei are in the
final stages of merging, similar to conclusions drawn from CO studies
(SSY99), but instead of being coplanar with the main CO disk (in which
the eastern nucleus is embedded), we suggest that the western nucleus
lies above it and will soon complete its final merger with the main
disk. The bridge of \HI\ seen connecting the two nuclei further supports
this scenario.  We suggest that the collection of radio supernovae
(RSN), detected in VLBA studies, in the more compact western nucleus
represent the second burst of star formation associated with this
final merger stage and that free-free absorption due to ionised gas in
the bulge-like component can account for the observed RSN
distribution.

\section{Acknowledgements}
We thank Toby Moore, Nick Scoville, John Porter, Andrew Wilson,
Sylvain Veilleux and Lee Mundy for useful discussions and Peter
Thomasson for help with the MERLIN data. We are grateful to Willem
Baan for kindly providing MERLIN 5-GHz-derived positions prior to
publication and the anonymous referee for useful comments. CGM
acknowledges financial support from The Royal Society.  This research
was partially supported by NSF grant AST9527289 to the University of
Maryland.  MERLIN is a U.K. national facility operated by the
University of Manchester on behalf of the Particle Physics and
Astronomy Research Council. This research has made use of NASA's
Astrophysics Data System Abstract Service (ADS) and the NASA/IPAC
Extragalactic Database (NED), which is operated by the Jet Propulsion
Laboratory, California Institute of Technology, under contract with
the National Aeronautics and Space Administration.

\clearpage

\figcaption[]{MERLIN 1.4-GHz radio continuum image of the central $\sim$ 900 pc of
   Arp~220. The beam size, indicated by a circle on the lower left
 corner, is
   0\farcs22. The two nuclei are labelled E (east) and W (west),
with the labels
   (A, B, C) assigned by BH95 given in
 parentheses. The weak,
   extended spur of emission, to the north-west
 of W, is labelled T. The contour levels, in multiples of
 3~$\times$~rms, are (-1, 1, 2, 4,
   8, 16, 32, 64,
 128)~$\times$~0.33~mJy~beam$^{-1}$. \label{cont}}

\figcaption[]{Channel maps of \HI\ absorption in Arp~220. The 1.4-GHz radio
 continuum
   image is shown the the top left panel with the beamsize,
 shown as a circle, in the
   lower left corner of this image. The
 individual channel maps, in which \HI\
   absorption is detected, are
 labelled according to the central velocity of each
   channel - indicated
 in the top right corner of each line image; the (dotted)
   contour
 levels in the absorption channel maps are (-64, -32, -16, -8, -4, -2,
 -1,
   1)~$\times$~1.5 \mJyb.
   \label{chanmaps}}

\figcaption[]{{\bf Central image~:} 1.4-GHz radio continuum map with the
 contours of
   the \HI\ absorption velocity field superimposed (contours
 from 5300 to 5600~\kms,
   in steps of 50~\kms).  {\bf Profiles~:}
 1.4-GHz radio continumm and \HI\
   absorption line velocity profiles at
 eight selected locations. Properties of the
   radio continuum and \HI\
 absorption at these locations are given in
   Tab.~\ref{TableResultsFit}.
   \label{imagespecs}}

\figcaption[]{{\bf Left panel~:} 1.4-GHz radio continuum map with the
   contours of the \HI\ absorption velocity field superimposed
   (contours from 5300 to 5600~\kms, in steps of 50~\kms).  {\bf
   Middle panel~:} Map of the equivalent hydrogen column density for
   the main \HI\ absorption component (assuming a spin temperature of
   100~K), as derived by multiple Gaussian fitting of the \HI\
   absorption line (see Sect.~\ref{SectionObservationsAnalysis}).
   Only regions where the 1.4-GHz continuum emission and the depth of
   the absorption line were stronger than 1.5~\mJyb\ have been mapped
   (i.e. 3 times the typical rms of the noise in the spectra). The
   contours are those of the velocity field of the main \HI\
   absorption component (contours from 5300 to 5600~\kms, in steps of
   50~\kms), derived from the same fit. {\bf Right panel~:} Velocity
   field of the main \HI\ absorption component, as derived by multiple
   Gaussian fitting of the \HI\ absorption line (see
   Sect.~\ref{SectionObservationsAnalysis}). Again, we have only
   mapped regions where the 1.4-GHz continuum emission and the depth
   of the absorption line were stronger than 1.5~\mJyb\ (i.e. 3 times
   the typical rms of the noise in the spectra). The contours are
   those of the 1.4-GHz radio continuum map (dotted contours~: 2, 4, 6
   and 8~\mJyb; solid contours~: 10, 20, 30, 40, 50 and 60~\mJyb).
   North is up and east is left throughout.  \label{FigureResultsFit}}

 \figcaption[]{Velocity profiles of the \HI\ absorption line at three
selected locations
   of the eastern nucleus (see
 Fig.~\protect\ref{imagespecs} and
   Tab.~\protect\ref{TableResultsFit}), with the fitted profiles
 superimposed
   (single component Gaussian fit~: dashed line; two
 component Gaussian fit~: dotted
   line, central panel only). The upper
 part of each panel displays the residuals of
   the subtraction of the
 fitted spectrum to the observed one (single component
   Gaussian fit~:
 solid line; two component Gaussian fit~: dotted line, central
   panel
 only). \label{FigureResultsSpectraE}}

\figcaption[]{1.4-GHz radio continuum and velocity profiles of the \HI\
absorption line
   at three selected locations of the western nucleus
 (see
   Fig.~\protect\ref{imagespecs} and
 Table~\protect\ref{TableResultsFit}),
   with the fitted profiles
 superimposed (single component Gaussian fit~: dashed
   line; multiple
 component Gaussian fit~: dotted line). The upper part of each
   panel
 displays the residuals of the subtraction of the fitted spectrum to
 the
   observed one (single component Gaussian fit~: solid line; multiple
 component
   Gaussian fit~: dotted line).
   \label{FigureResultsSpectraW}}

\figcaption[]{Sketch of merging geometry of prograde western disk with
(retrograde) eastern/main disk, as viewed on the sky. Small arrows
indicate rotation direction of each disk; the orbital/merger path of
the western disk is also indicated.
\label{mergesketch}}

\figcaption[]{1.4-GHz MERLIN radio continuum of Arp~220 with the
   locations of radio supernovae (RSN) and OH masers (from SLLD98)
   indicated. The position of the E and W nuclei as derived from
   MERLIN radio continuum observations at 1.4 GHz (present paper) and
   5 GHz (Baan, private communication) are also shown.
   \label{MERLINpluscrosses}}

\begin{figure}
\plotone{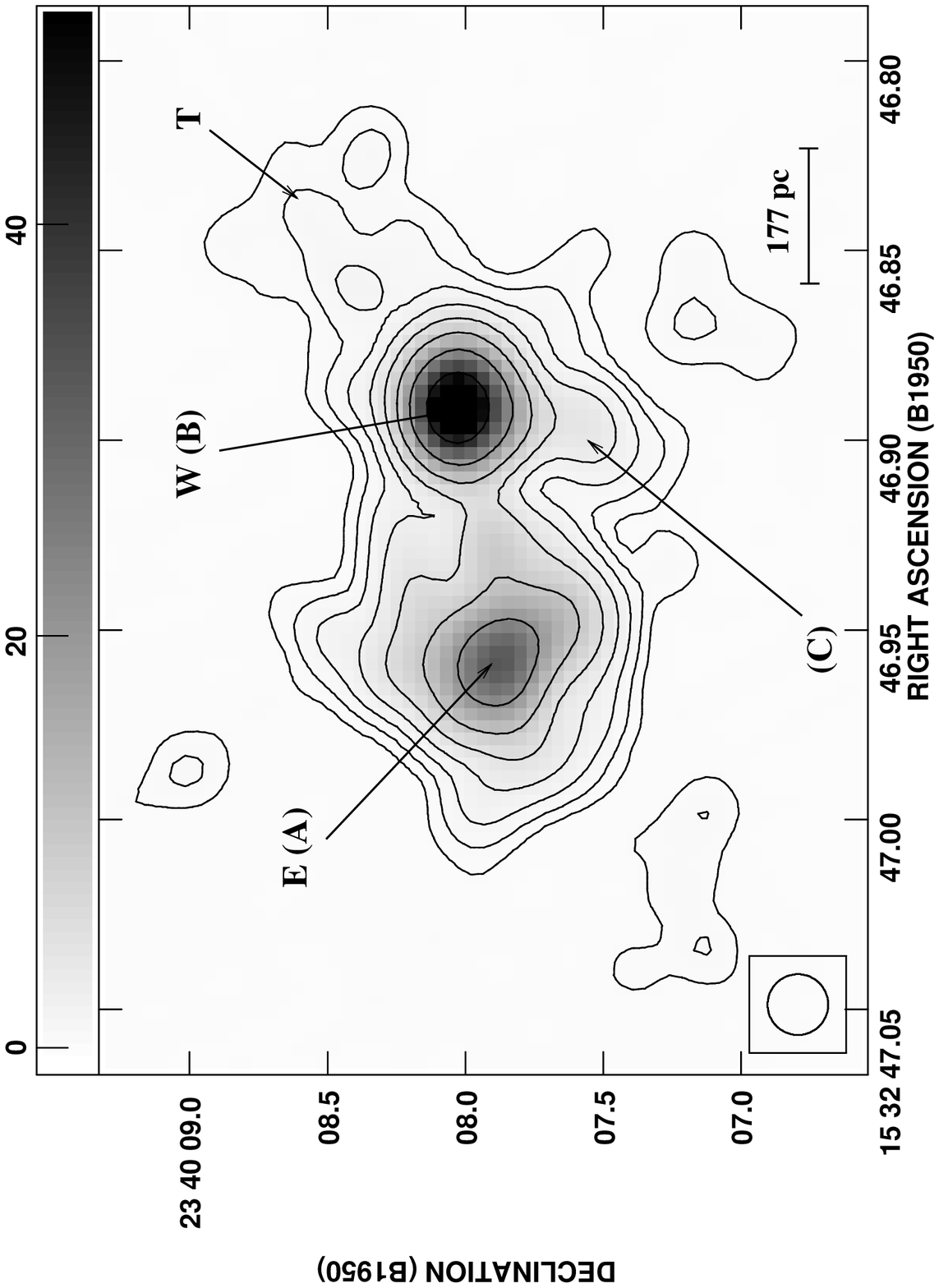}{Figure~1}
\end{figure}

\plotone{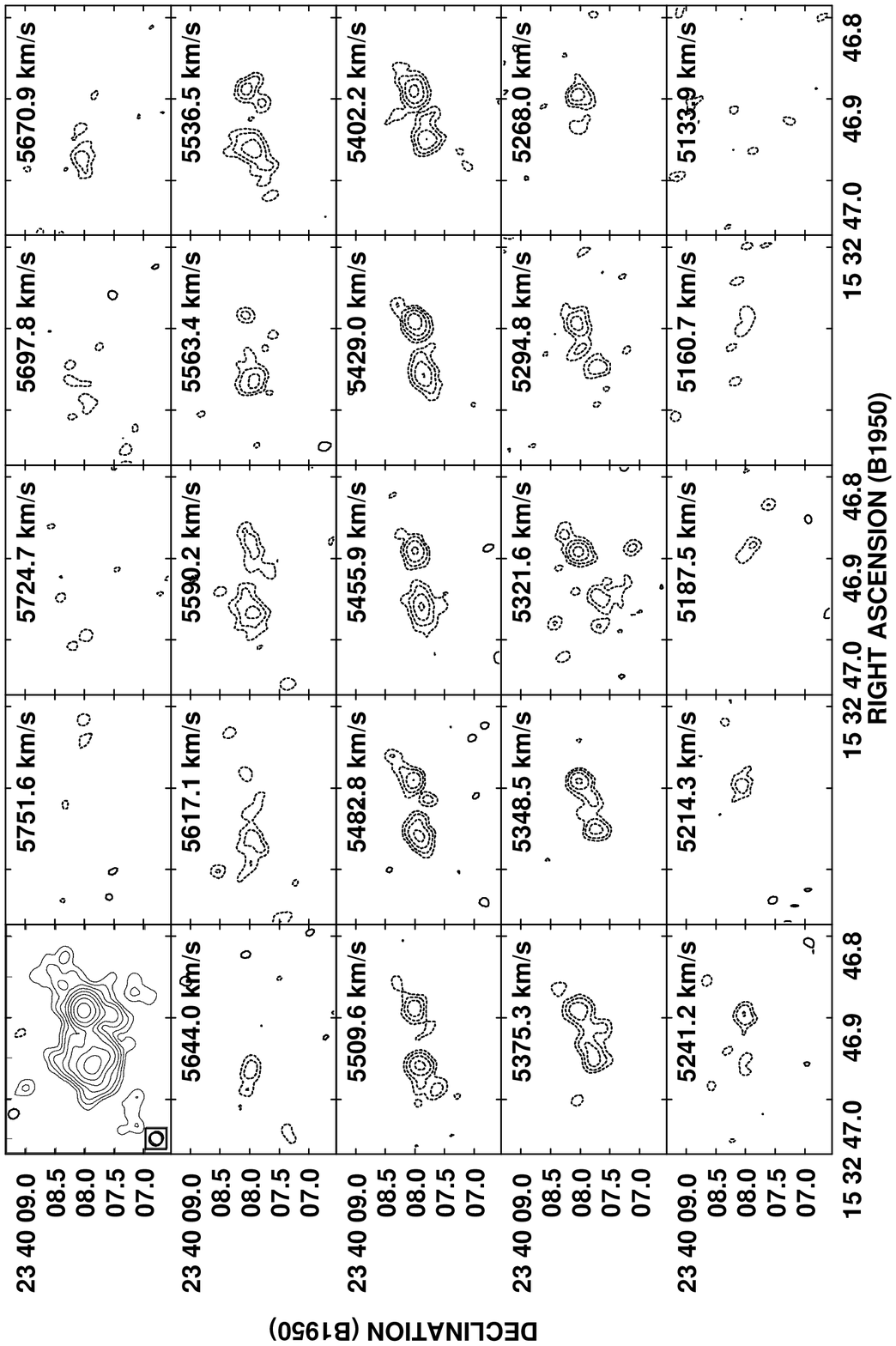}{Figure~2}

\plotone{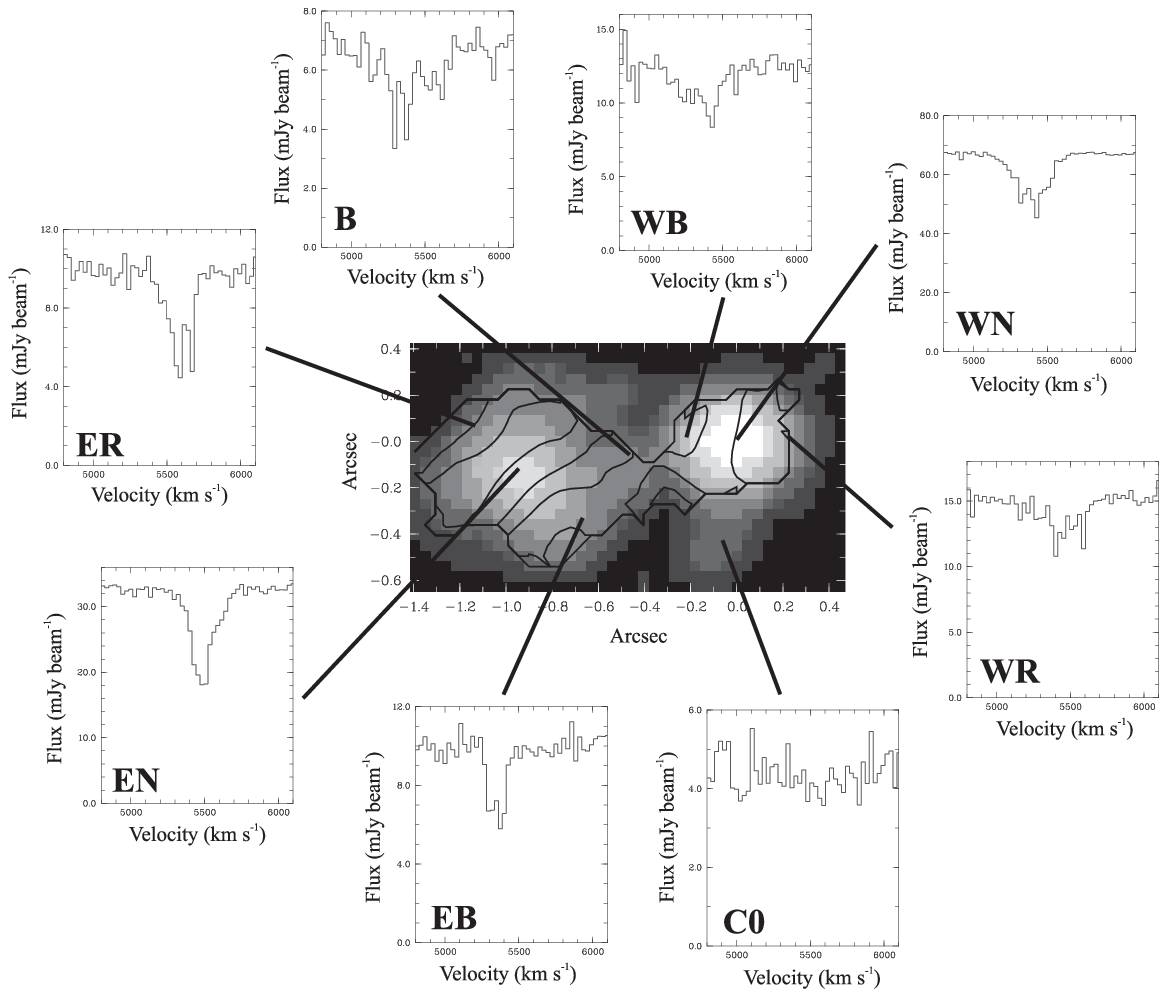}{Figure~3}

\plotone{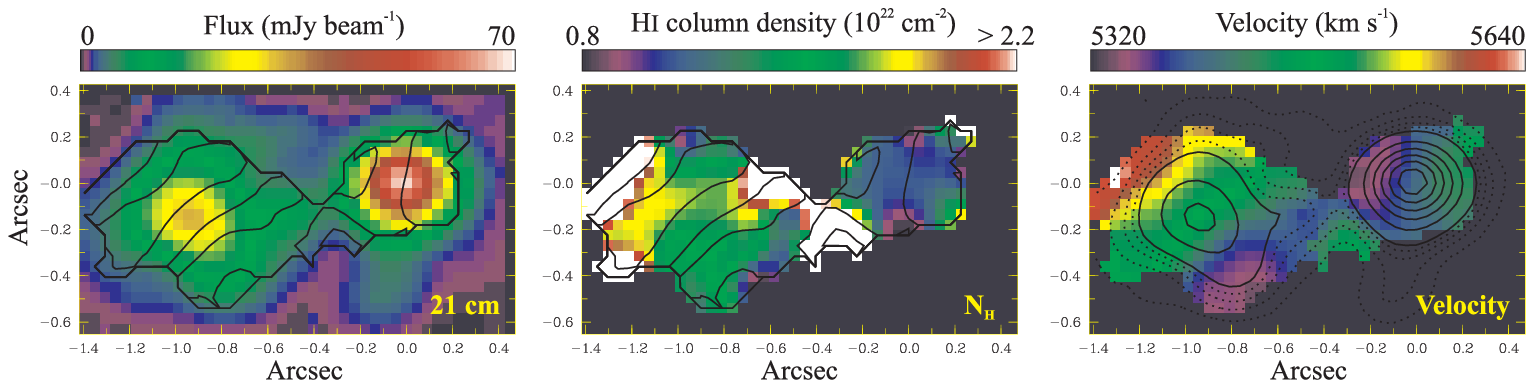}{Figure~4}

\plotone{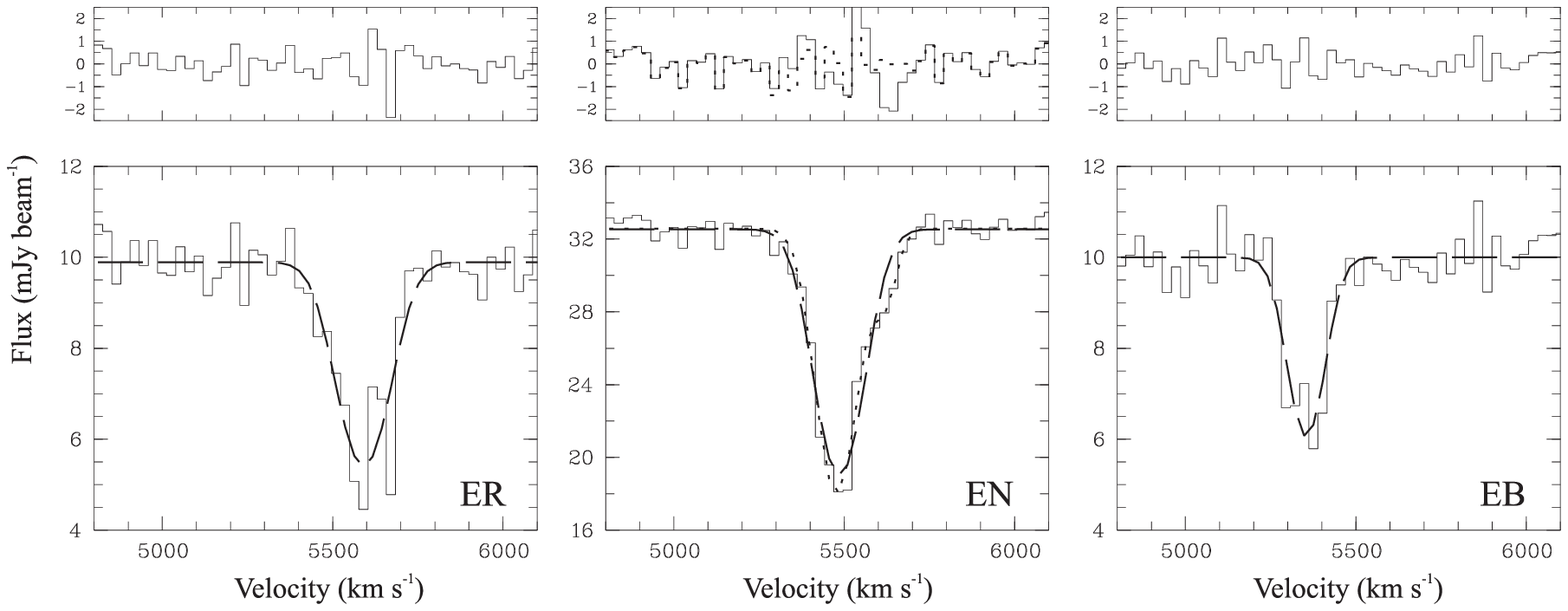}{Figure~5}
\plotone{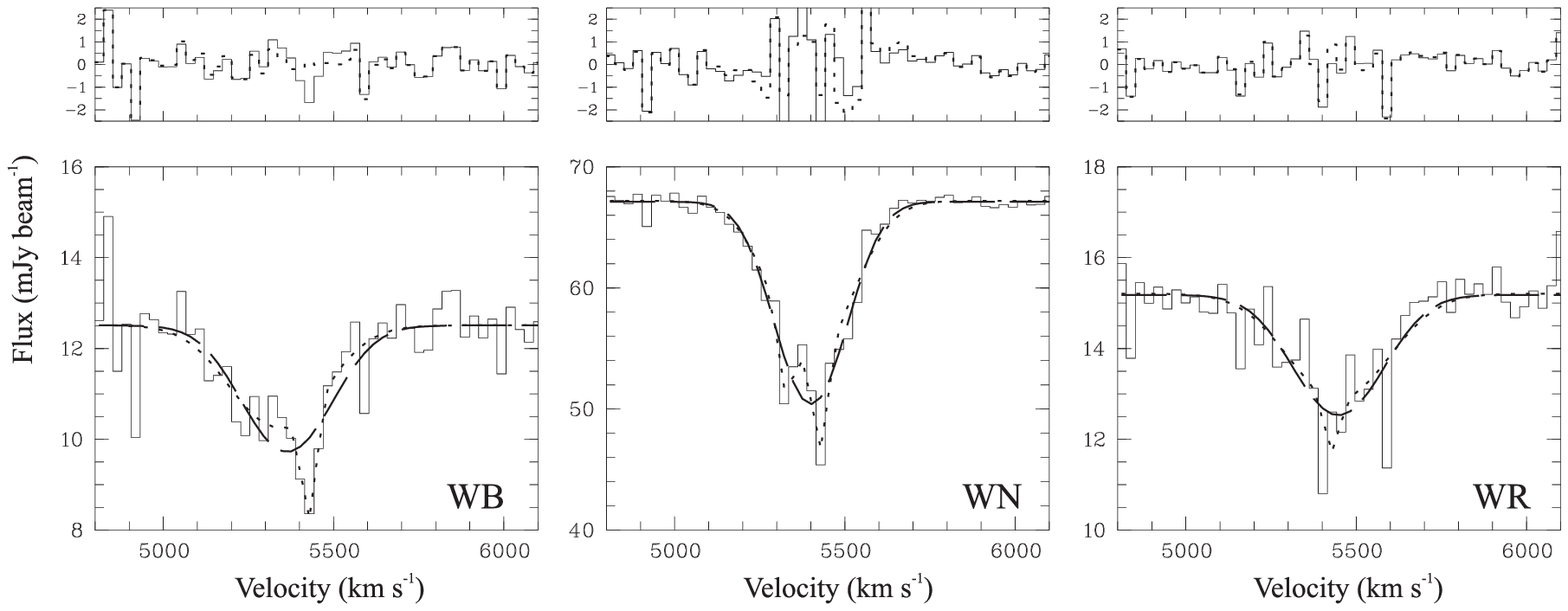}{Figure~6}
\plotone{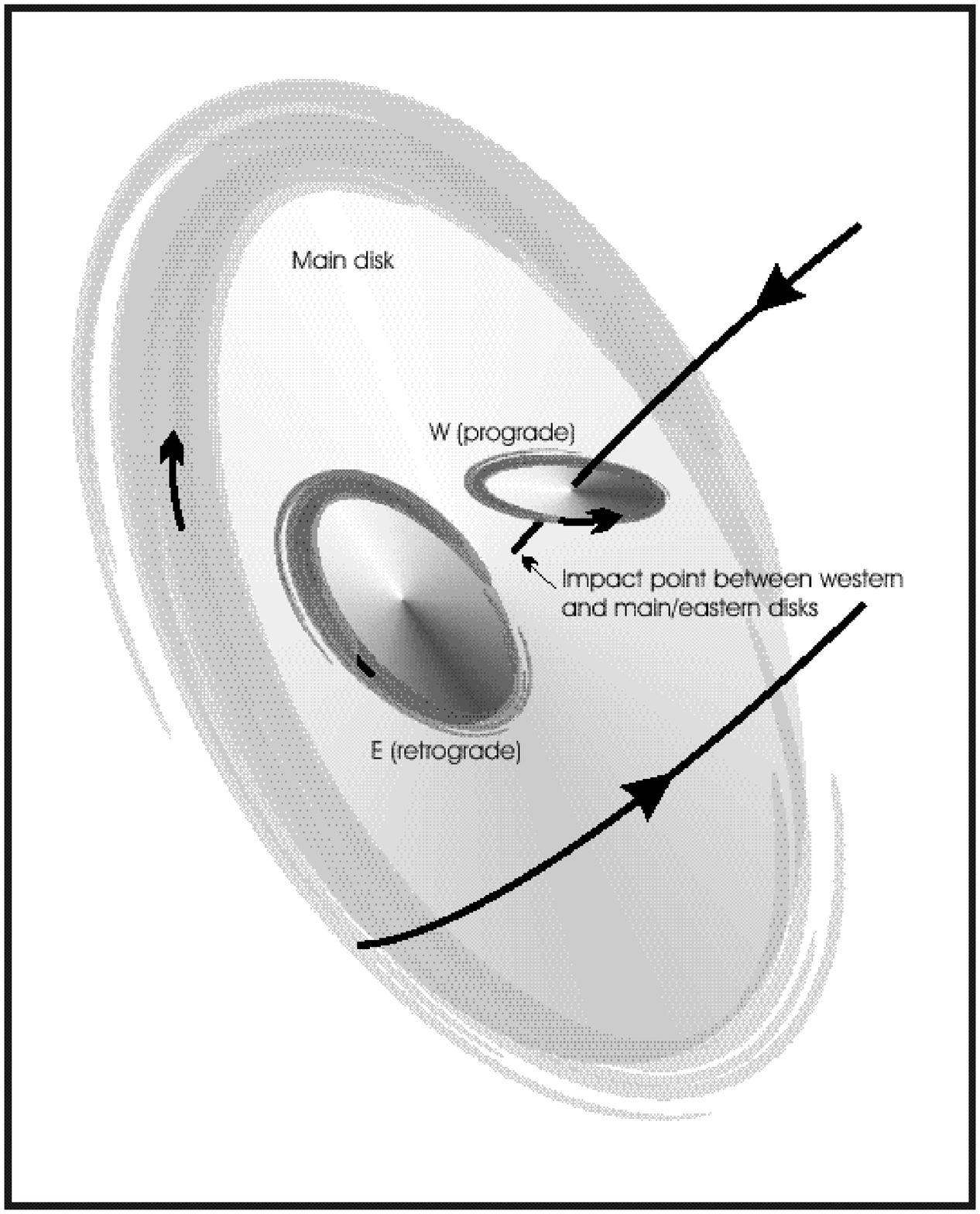}{Figure~7}
\plotone{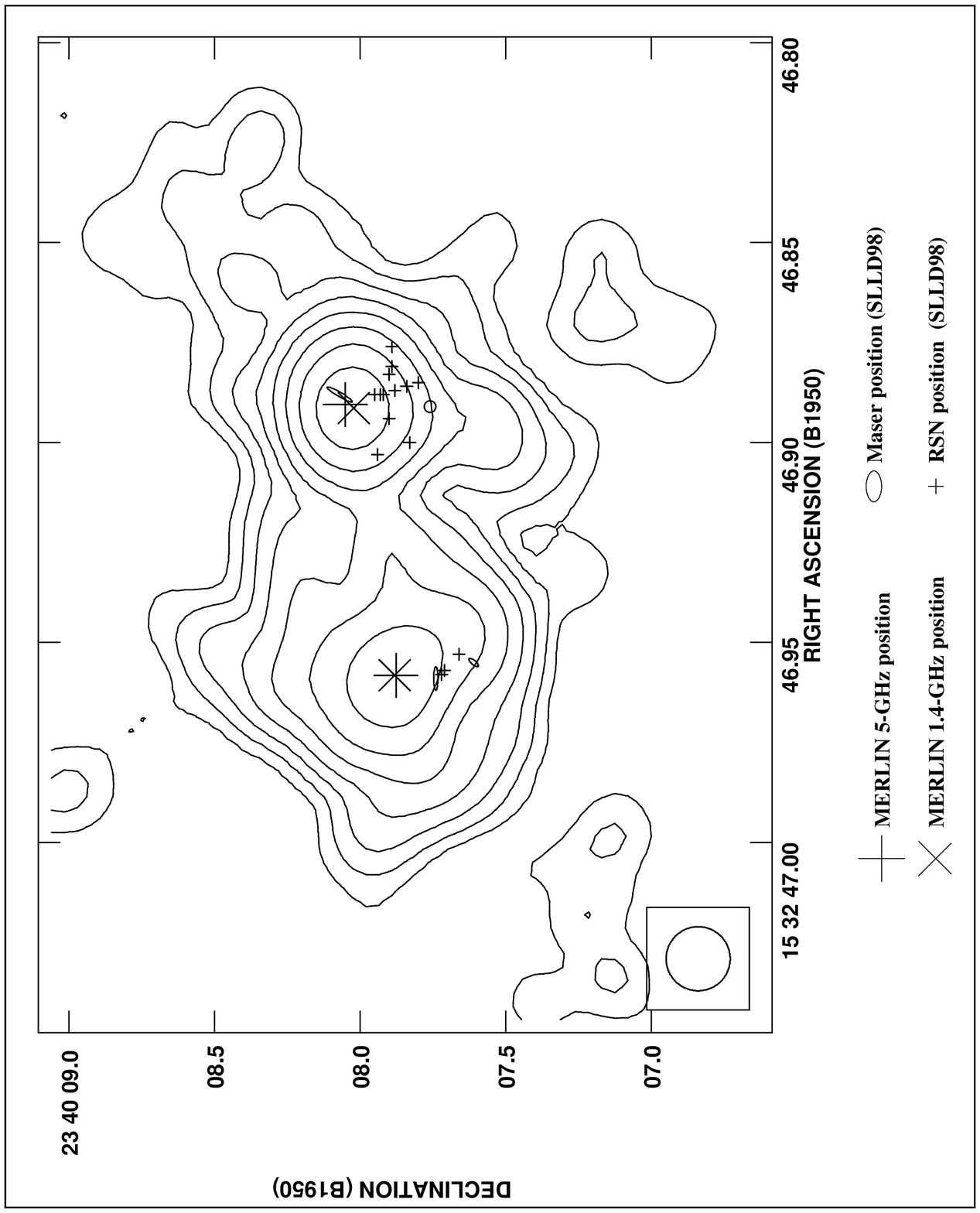}{Figure~8}

\clearpage

\setcounter{table}{0} 
\begin{table}
   \caption[]{Results of single and multiple component Gaussian
   fitting of spectra at seven selected locations where \HI\
   absorption is detected (see Fig.~\protect\ref{imagespecs}). The
   offset ($\Delta\alpha$,$\Delta\delta$) of each location is given
   relative to WN, where the derived position for WN is $\alpha_{\rm
   1950}$ = 15$^h$32$^m$46\fsec89142s, $\delta_{\rm 1950}$ =
   23\arcdeg40\arcmin 08\farcs0215. For each spectral component the
   measured 1.4-GHz continuum flux density (F$_{\rm 1.4 GHz}$) is
   given along with derived \HI\  column density (N$_{\rm H}$/T$_{\rm
   S}$) and corresponding visual extinction (A$_{\rm V}$/T$_{\rm S}$)
   for a spin temperature of T$_{\rm S}$, absorption line velocity
   centroid and full width at half maximum (FWHM). Details on the
   Gaussian fitting procedure can be found in
   Sect.~\ref{SectionObservationsAnalysis}.
\label{TableResultsFit}}
\scriptsize
   \begin{tabular}{lccccccc}
      \noalign{\smallskip}\hline\hline\noalign{\smallskip}
      & B & ER & EN & EB & WB & WN & WR \\
      \noalign{\smallskip}\hline\noalign{\smallskip}
      $\Delta\alpha$ (arcsec) & $-$0.405 & $-$1.125 & $-$0.945 & $-$0.675
      & $-$0.270 & 0 & $+$0.225\\
      $\Delta\delta$ (arcsec) & $-$0.090 & $+$0.045 & $-$0.135 & $-$0.360
      & 0 & 0 & 0 \\
      \noalign{\smallskip}\hline\noalign{\smallskip}
      \multicolumn{8}{l}{\bf Single component Gaussian fitting}\\
      F$_{1.4\,\mbox{\tiny GHz}}$ (\mJyb) & 6.86$\pm$0.39 & 9.89$\pm$0.52 & 
      32.53$\pm$1.63& 10.00$\pm$0.52 & 12.51$\pm$0.63 & 67.12$\pm$3.36 & 
      15.18$\pm$0.77 \\
      N$_{\mbox{\tiny H}}$/T$_{\rm s}$ (10$^{20}$~cm$^{-2}$) & 2.4$\pm$0.4 & 1.99$\pm$0.15 & 
      1.68$\pm$0.04 & 1.17$\pm$0.11 & 1.46$\pm$0.16 & 1.32$\pm$0.02 & 
      1.13$\pm$0.15 \\
      A$_{\rm V}$/T$_{\rm s}$ (10$^{-2}$mag) & 15.5$\pm$2.6
      &12.9$\pm$1.0 &10.9$\pm$0.3 & 7.6$\pm$0.7 &
      9.4$\pm$1.0 & 8.5$\pm$0.1 & 7.3$\pm$1.0 \\
      Velocity (\kms) & 5387$\pm$20 & 5591$\pm$5 & 5486$\pm$19 & 5354$\pm$5 & 
      5364$\pm$13 & 5401$\pm$2 & 5448$\pm$12 \\
      FWHM (\kms) & 390$\pm$70 & 169$\pm$12 & 161$\pm$5 & 120$\pm$12 & 
      300$\pm$40 & 238$\pm$5 & 300$\pm$40 \\
      \noalign{\smallskip}\hline\noalign{\smallskip}
      \multicolumn{8}{l}{\bf Multiple component Gaussian fitting}\\
      F$_{1.4\,\mbox{\tiny GHz}}$ (\mJyb) & -- & -- & 32.57$\pm$0.1.63 & -- 
      & 12.51$\pm$0.64 & 67.19$\pm$3.36 & 15.20$\pm$0.78 \\
      \multicolumn{8}{l}{$\triangleright$Main component properties}\\
      N$_{\mbox{\tiny H}}$/T$_{\rm s}$ (10$^{20}$~cm$^{-2}$) & -- & -- & 1.52$\pm$0.05 & -- 
      & 1.22$\pm$0.18 & 1.10$\pm$0.03 & 1.08$\pm$0.15 \\
      A$_{\rm V}$/T$_{\rm s}$ (10$^{-2}$mag)& -- & -- &9.8$\pm$0.3 & -- 
      & 7.9$\pm$1.2 & 7.1$\pm$0.2 & 7.0$\pm$1.0 \\
      Velocity (\kms) & -- & -- & 5478$\pm$2 & -- 
      & 5336$\pm$18 & 5404$\pm$3 & 5449$\pm$16 \\
      FWHM (\kms) & -- & -- & 132$\pm$5 & -- 
      & 317$\pm$50 & 270$\pm$7 & 340$\pm$50 \\
      \multicolumn{8}{l}{$\triangleright$Sub-component properties}\\
      N$_{\mbox{\tiny H}}$/T$_{\rm s}$ (10$^{20}$~cm$^{-2}$) & -- & -- & 0.20$\pm$0.03 & -- 
      & $<$ 0.1 & 0.10$\pm$0.01 & $<$ 0.08 \\
      A$_{\rm V}$/T$_{\rm s}$ (10$^{-2}$mag) & -- & -- & 1.3$\pm$0.2 & -- 
      & -- & 0.6$\pm$0.1 & --\\
      Velocity (\kms) & -- & -- & 5618$\pm$6 & -- 
      & 5326$^a$ & 5326$^a$ & 5326$^a$ \\
      FWHM (\kms) & -- & -- & 80$\pm$12 & -- 
      & 50$^a$ & 50$^a$ & 50$^a$ \\
      N$_{\mbox{\tiny H}}$/T$_{\rm s}$ (10$^{20}$~cm$^{-2}$) & -- & -- & -- & -- 
      & 0.26$\pm$0.05 & 0.15$\pm$0.02 & 0.10$\pm$0.04 \\
      A$_{\rm V}$/T$_{\rm s}$ (10$^{-2}$mag) & -- & -- & -- & -- 
      & 1.7$\pm$0.3 & 1.0$\pm$0.1 & 0.6$\pm$0.3 \\
      Velocity (\kms) & -- & -- & -- & -- 
      & 5428$^a$ & 5428$^a$ & 5428$^a$ \\
      FWHM (\kms) & -- & -- & -- & -- 
      & 50$^a$ & 50$^a$ & 50$^a$ \\
      \noalign{\smallskip}\hline\noalign{\smallskip}
      \multicolumn{7}{l}{$^a$~Parameter set to a fixed value during the fit.}\\
   \end{tabular}
\end{table}


\end{document}